\documentclass[preprint, showpacs,preprintnumbers,aps,prd,amsmath,amssymb]{revtex4}

\usepackage{graphicx}
\usepackage{dcolumn} 
\usepackage{bm}      

\newcommand{\bea}{\begin{eqnarray}}
\newcommand{\eea}{\end{eqnarray}}

\begin{document}
\title{Can universe exit from phantom inflation \\ due to gravitational back reaction? }         
\author{  Puxun Wu and Hongwei Yu
\footnote{To whom correspondence should be addressed} }

\affiliation { Department of Physics and Institute of  Physics,\\
Hunan Normal University, Changsha, Hunan 410081, China}


\begin{abstract}
The effects of the gravitational back reaction of cosmological
perturbations are investigated in a phantom inflation model. The
effective energy-momentum tensor of the gravitational back
reaction of cosmological perturbations whose wavelengths are
larger than the Hubble radius is calculated. Our results show that
the effects of gravitational back reaction will counteract that of
the phantom energy. It is demonstrated in a chaotic phantom
inflation model that if the phantom field at the end of inflation
is larger than a critical value determined by the necessary
e-folds, the phantom inflation phase might be terminated by the
gravitational back reaction.
\end{abstract}


 \maketitle

\baselineskip=14pt

\section{Introduction}
A phantom field~\cite{Cal}, which is motivated by recent
observations, appears as a possible candidate of dark energy to
explain the present acceleration of cosmic expansion  and has
received increasing attention~\cite{Phan1, Phan2, Phan3, Carr, Cald,
WuYu, Piao, PFG, Nojiri, Vikman, Sami}. It has an unusual kinetic
term in its Lagrangian which leads to the super-negative equation of
state($ w \equiv p/\rho\leq-1$) and gives rise to some strange
properties, such as the violation of the dominant energy
condition~\cite{Carr}, the catastrophic quantum instability of
vacuum with respect to  its decay into gravitons and phantom
particles with negative energy~\cite{Carr},   the increase of energy
density with time and the big rip~\cite{Cald}.

Recently authors in Refs.~\cite{Piao, PFG} considered a scenario
in which the primordial inflation is driven by phantom matter.
Since the phantom field climbs up along its potential due to the
negative kinetic term in contrast to the normal scalar field
rolling down the potential, in phantom inflation the slow climb
parameters replace the general slow roll ones in normal inflation
models. During the slow climb expansion the nearly scale invariant
spectrum can be generated. However, there exists a problem as to
whether the universe can smoothly exit from the phantom
inflationary phase. If the universe enters the phantom
inflationary phase, it seems to continue up to  big rip after some
time or to a de Sitter expansion forever, depending on the
steepness of the phantom potential~\cite{Sami}. Furthermore, there
exists a no-go theorem concerning the graceful exit of single
phantom field inflation, which states that the dynamical
transition from states with $ w<-1$ to those with $ w>-1$ are
physically implausible~\cite{Vikman}. So far, some mechanisms
involving new physics, such as adding an additional normal scalar
field into the phantom potential~\cite{Piao}, "Big
Trip"~\cite{PFG}, varying the equation of state~\cite{Nojiri}, are
proposed to solve this problem of graceful exit. It is worth
noting that the exit problem may also be resolved in the inflation
with higher derivatives~\cite{Vik}.

In the present paper, we would like to examine the back reaction of
cosmological perturbations in  the phantom inflation scenario and
to see if the problem of exit may be resolved in the phantom
inflation without adding new physics. Let us note that the
gravitational back reaction of the normal scalar field inflation has
been studied in~\cite{Raul} and it was  found that the back reaction
may lead to a shortening of the period of inflation and becomes very
important before the end of inflation if the initial value of
inflation scalar field is larger than a critical value.

\section{phantom inflation}  
In this section we will simply introduce the phantom inflation
model. A general  effective action of a phantom field can be
expressed as
\begin{eqnarray}\label{lag}
{\cal L}_{phan}=-\frac{1}{2}(\partial_\mu \varphi)^2-V(\varphi),
\end{eqnarray}
where $V(\varphi)$ is the potential of the phantom field and the
metric signature is $(+ - - -)$. Thus we can obtain the energy
momentum tensor
\begin{eqnarray}\label{EMT}
T_{\mu\nu}=-\partial_\mu \varphi \partial_\nu \varphi
+g_{\mu\nu}\left[\;
\frac{1}{2}\partial^\alpha\varphi\partial_\alpha\varphi+V(\varphi)\right]\;,
\end{eqnarray}
and the energy density $\rho$ and  pressure $p$
\begin{eqnarray}
\rho=-\frac{1}{2}\dot{\varphi}^2+V(\varphi),\qquad
p=-\frac{1}{2}\dot{\varphi}^2-V(\varphi)\;,
\end{eqnarray}
where a dot means derivative with respect to time. Clearly, one
has that $\omega\equiv p/\rho<-1$.

Assuming  the phantom field is minimally coupled with gravity and
the early universe is dominated by phantom matter, the Friedmann
equation for a flat geometry with scale factor $a(t)$ can be
expressed as
\begin{eqnarray}\label{Fd}
H^2\equiv\bigg(\frac{\dot{a}}{a}\bigg)^2=\frac{8\pi}
{3m_{pl}^2}\bigg(-\frac{1}{2}\dot{\varphi}^2+V(\varphi)\bigg)\;,
\end{eqnarray}
and the phantom field satisfies the following dynamical equation
\begin{eqnarray}\label{pheq}
\ddot{\varphi}+3H\dot{\varphi}-V'(\varphi)=0\;.
\end{eqnarray}
Here $m_{pl}$ is the Planck mass and a prime denotes derivative
with respect to field $\varphi$. Obviously the Eq.~(\ref{Fd},
\ref{pheq}) are different from the equations in the normal
inflation models with the minus signs replacing the plus  before
the kinetic and $V'(\varphi)$ terms.  In order to implement the
inflation the slow-climb parameters should be considered, which
are defined as~\cite{Piao}
\begin{eqnarray}
\epsilon_{pha}\equiv -\frac{\dot{H}}{H^2}\;, \qquad
\delta_{pha}\equiv-\frac{\ddot{\varphi}}{\dot{\varphi}H}\;.
\end{eqnarray}
If the conditions $|\epsilon_{pha}|\ll 1$ and $|\delta_{pha}|\ll
1$ are satisfied, we can simplify Eqs.(\ref{Fd}) and (\ref{pheq})
\begin{eqnarray}\label{Ifd}
H^2\simeq\frac{8\pi }{3m_{pl}}V(\varphi)\;,
\end{eqnarray}
\begin{eqnarray}\label{Ipe}
3H\dot{\varphi}-V'(\varphi)\simeq 0\;.
\end{eqnarray}
 It follows that
\begin{eqnarray}\label{ava}
a(\varphi)\sim \exp\left[\frac{8\pi}{m_{pl}^2}\int
\frac{V}{V'}d\varphi\right]\;.
\end{eqnarray}
Therefore, approximately, the universe expands exponentially. If the
universe enters the phantom inflationary phase, it will continue up
to big rip after some time
or to a de Sitter expansion.
So there exists a problem of graceful
exit from the inflationary era to the radiation dominated one
\cite{Piao, PFG}. In what follows,  we will discuss the
gravitational back reaction of cosmological perturbations in phantom
inflation to see if this seemingly ever-going inflation could be
terminated.

\section{Gravitational back reaction}
Due to non-linearity of Einstein equation, the cosmological
perturbations back react on the background  metric.  This back
reaction effect on the background  can be  characterized by a
gauge-invariant effective energy-momentum tensor
$\tau_{\mu\nu}$~\cite{Raul}
\begin{eqnarray}\label{1}
\tau_{\mu\nu}=\langle T_{\mu\nu}^{(2)}-\frac{m_{pl}^2}{8\pi
}{G_{\mu\nu}^{(2)}}\rangle,
\end{eqnarray}
where $T_{\mu\nu}^{(2)}$ and $G_{\mu\nu}^{(2)}$ express the second
order metric and matter perturbations and pointed brackets stand
for spatial averaging. This formalism can be applied  to both
scalar and tensor perturbations and applies independent of the
wavelength of the perturbations. The effects of gravitational back
reactions  have been studied in the context of cosmological
models, for example, they have been used to address the issue of
dynamical relaxation of the cosmological constant~\cite{Bran}, the
possible termination of quintessence phase~\cite{Li} and the
avoidance of big rip in phantom cosmology~\cite{WuYu}.

In inflationary era, the phase space of infrared modes grows
rapidly since the wavelengths  are stretched exponentially while
the Hubble radius is nearly constant. Hence we expect the back
reaction effect of infrared modes  grows and plays the main role.
So we will only consider the effect of infrared modes on the
evolution of the universe.
In longitudinal gauge, the  metric with
scalar perturbations can be written as
\begin{eqnarray}\label{3}
ds^2=(1+2\Phi)dt^2-a(t)^2(1-2\Phi)\delta_{ij}dx^i dx^j\;,
\end{eqnarray}
where $\Phi$ is the Bardeen potential. In addition to the
geometrical perturbations, the perturbations of the phantom field,
$\delta\varphi$, during the inflation must also be considered.
 Expansion of the energy-momentum tensor $T_{\mu\nu}$ and
Einstein tensor $G_{\mu\nu}$  to the second order in $\Phi$ and
$\delta\varphi$ yields the non vanishing components of the
effective back reaction energy-momentum tensor
\begin{eqnarray}\label{t00}
\tau_{00}&=&\frac{m_{pl}^2}{8\pi}
[12H\langle\Phi\dot{\Phi}\rangle-3\langle\dot{\Phi}^2\rangle+9a^{-2}\langle\nabla
\Phi^2\rangle] -\frac{1}{2}\langle\delta\dot{\varphi}^2\rangle
\nonumber\\
&&\;-\frac{1}{2}a^{-2}\langle(\nabla\delta\varphi)^2\rangle
+\frac{1}{2}V,_{\varphi\varphi}\langle\delta\varphi^2\rangle+2V,_\varphi\langle\Phi\delta\varphi\rangle\;,
\end{eqnarray}
and
\begin{eqnarray}\label{tij}
\tau_{ij}&=&a^2 \delta_{ij}{\biggl\{}\frac{m_{pl}^2}{8\pi}{\biggl
[}(24H^2+16
\dot{H})\langle\Phi^2\rangle+24H\langle\Phi\dot{\Phi}\rangle
 +\langle\dot{\Phi}^2\rangle+4\langle\Phi\ddot{\Phi}\rangle-
\frac{3}{4}a^{-2}\langle\nabla \Phi^2\rangle{\biggl ]}\nonumber\\
&&
-4\dot{\varphi}_0^2\langle\Phi^2\rangle-\frac{1}{2}\langle\delta\dot{\varphi}^2\rangle
-\frac{1}{2}a^{-2}\langle(\nabla\delta\varphi)^2\rangle+4\dot{\varphi}_0\langle\Phi\delta\dot{\varphi}\rangle
\nonumber\\
&&-\frac{1}{2}V,_{\varphi\varphi}\langle\delta\varphi^2\rangle+2V,_\varphi\langle\Phi\delta\varphi\rangle{\biggl\}}\;.
\end{eqnarray}
 Although the linear perturbation equations are different from that in  which matter perturbations are induced by
 a normal scalar field,
  combining these linear equations we obtain the second order partial
differential equation for $\Phi$ in the phantom dominated universe
 which is the same as that in  normal scalar field dominated
 one~\cite{WuYu}.
Thus for long-wavelength perturbations, we have \cite{Mukh}
\begin{eqnarray}\label{Phk}
\Phi_k\simeq -\frac{A_km_{pl}^2}{16\pi}\frac{V'^2}{V^2}\;.
\end{eqnarray}
Here $A_k$ is an integration constant.  Applying  the slow climb
conditions, we can easily obtain
\begin{eqnarray}\label{dPh}
\dot{\Phi}_k\simeq
-\frac{2V}{3H}\bigg[\frac{V''}{V}-\frac{V'^2}{V^2}\bigg]\Phi_k\;,
\end{eqnarray}
and
\begin{eqnarray}\label{ddPh}
\ddot{\Phi}_k\simeq \frac{m_{pl}^2}{24\pi
}\bigg[4\frac{V''^2}{V}-2\frac{V'V'''}{V}
-3\frac{V''V'^2}{V^2}+\frac{V'^4}{V^3}\bigg]\Phi_k\;.
\end{eqnarray}
 Expanding the Einstein equation to first order, the $0i$
equation gives rise to a constraint relating $\Phi$ and
$\delta\varphi$:
\begin{eqnarray}\label{E0i}
\dot{\Phi}+H\Phi=-\frac{4\pi}{m_{pl}^2}\dot{\varphi}_0\delta\varphi\;.
\end{eqnarray}
Using the above expression we find
\begin{eqnarray}\label{del}
\delta\varphi\simeq\left[-2\frac{V}{V'}+\frac{m_{pl}^2}{2\pi
}\left(\frac{V''}{V'}-\frac{V'}{V}\right)\right]\Phi\;,
\end{eqnarray}
and
\begin{eqnarray}\label{ddel}
\delta\dot{\varphi}\simeq\frac{V'}{3H}\left[-2+2\frac{V
V''}{V'^2}+4\frac{V''}{V}-4\left(\frac{V'}{V}\right)^2+\frac{m_{pl}^2}{2\pi
}\left(\frac{V'''}{V'}-3\frac{V''^2}{V'^2}+3\frac{V''}{V}-\frac{V'^2}{V^2}\right)\right]
\Phi\;.
\end{eqnarray}
Substituting Eqs.~(\ref{dPh}, \ref{ddPh}, \ref{del}, \ref{ddel})
into Eqs.~(\ref{t00}, \ref{tij}) and defining
$\rho_{br}\equiv\tau_0^0, p_{br}\equiv-\frac{1}{3}\tau_i^i$, we
get
\begin{eqnarray}\label{rhobr}
\rho_{br}&=&\bigg[2V\left(\frac{VV''}{V'^2}-2\right)+\frac{m_{pl}^2}{2\pi}\bigg(5V''
-\frac{3V'^2}{2V}-
\frac{7V''^2V}{2V'^2}\bigg)+\frac{m_{pl}^4}{2\pi^2}
\bigg(\frac{V''^3}{V'^2} \nonumber\\
&&-\frac{V'''V''}{4V'}- \frac{3V'^4}{8V^3}+\frac{3V''V'^2}{2V^2}
-\frac{17V''^2}{8V}+\frac{V'''V'}{4V}\bigg)\nonumber\\
&&-\frac{m_{pl}^6}{192\pi^3V}
\bigg(\frac{V'^3}{V^2}-\frac{3V''V'}{V}+\frac{3V''^2}{V'}
-V'''\bigg)^2\bigg]\langle\Phi^2\rangle\;,
\end{eqnarray}
and
\begin{eqnarray}\label{pbr}
p_{br}&=&
\bigg[2V\left(2-\frac{VV''}{V'^2}\right)+\frac{m_{pl}^2}{2\pi}\bigg(V''
-\frac{7V'^2}{6V}+
\frac{V''^2V}{2V'^2}\bigg)+\frac{m_{pl}^4}{2\pi^2}
\bigg(\frac{V''^3}{2V'^2}\nonumber\\
&&-\frac{V'''V''}{4V'}- \frac{V'^4}{3V^3}+ \frac{25V''V'^2}{24V^2}
-\frac{31V''^2}{24V}+\frac{V'''V'}{3V}\bigg)\nonumber\\
&&-\frac{m_{pl}^6}{192\pi^3V}
\bigg(\frac{V'^3}{V^2}-\frac{3V''V'}{V}+\frac{3V''^2}{V'}
-V'''\bigg)^2\bigg]\langle\Phi^2\rangle\;,
\end{eqnarray}
where all terms involving spatial gradients are dropped since only
the back reaction effect of infrared modes are considered. Note
that here the subscript ``$br$'' means back reaction. During the
phantom inflation the energy density of the background is
$\rho_{bg}\simeq V(\varphi)$, thus we have
\begin{eqnarray}\label{ratio}
\frac{\rho_{br}}{\rho_{bg}}&\simeq&
\bigg[2\left(\frac{VV''}{V'^2}-2\right)+\frac{m_{pl}^2}{2\pi}\bigg(\frac{5V''}{V}
-\frac{3V'^2}{2V^2}-
\frac{7V''^2}{2V'^2}\bigg)+\frac{m_{pl}^4}{2\pi^2}
\bigg(\frac{V''^3}{V'^2V}\nonumber\\
&&-\frac{V'''V''}{4V'V}- \frac{3V'^4}{8V^4} +\frac{3V''V'^2}{2V^3}
-\frac{17V''^2}{8V^2}+\frac{V'''V'}{4V^2}\bigg)\nonumber\\
&&-\frac{m_{pl}^6}{192\pi^3V^2}
\bigg(\frac{V'^3}{V^2}-\frac{3V''V'}{V}+\frac{3V''^2}{V'}
-V'''\bigg)^2\bigg]\langle\Phi^2\rangle\;.
\end{eqnarray}
If the above ratio is negative, the phantom energy will be
counteracted by the effects of the back reaction. When the ratio
becomes nearly negative unity, the phantom energy will be
counteracted completely and may become smaller than that of other
matter, such as gravitational waves generated during inflation or
some light scalar fields which can reheat the universe. As a
result, the universe will exit from the phantom inflationary phase
due to the gravitational
 back reaction and enters the radiation dominated era. In order to determine the
value of this ratio, it is crucial to evaluate the two-point
function $\langle \Phi^2\rangle$, which can be obtained by
integrating over all Fourier modes of $\Phi$:
\begin{eqnarray}
\langle\Phi^2\rangle=\int_{k_i}^{k_t}\frac{dk}{k}|\Phi_k^2|\;.
\end{eqnarray}
The integral runs over all modes with scales larger than the
Hubble radius $k<k_t=H(t)a(t)$ but smaller than the Hubble radius
at the initial time $t_i$ , i.e., $k>k_i=H(t_i)a(t_i)$, where
$t_i$ means the beginning time of phantom inflation. From the
differential equation of $\Phi$ given in Ref.~\cite{WuYu},   we
can  obtain
\begin{eqnarray}
\Phi_k=\frac{2}{\pi
m_{pl}^2}\left(\frac{\dot{\varphi_0}H^2}{\dot{H}}\right)_{t_H(k)}\left(
\frac{1}{a}\int adt\right)^. =\sqrt{\frac{1}{6\pi
m_{pl}^2}}\frac{V'^2}{V^2}\left(\frac{V^{3/2}}{V'}\right)_{t_H(k)}\;,
\end{eqnarray}
when $k_i<k<k_t$, where the index $t_H(k)$ means the time when the
scale $k$ crosses the Hubble radius. Thus we have
\begin{eqnarray}
\langle\Phi^2\rangle=\frac{1}{6\pi
m_{pl}^2}\frac{V'^4}{V^4}\int_{k_i}^{k_t}\left(\frac{V^3}{V'^2}\right)_{t_H}d\ln
k\;.
\end{eqnarray}
Using $d\ln k=d\ln a +d\ln H$ and neglecting $d\ln
H=-\epsilon_{pha}d\ln a$, we get
\begin{eqnarray}\label{Phiva}
\langle\Phi^2\rangle&=&\frac{1}{6\pi
m_{pl}^2}\frac{V'^4}{V^4}\int_{a_i}^{a_t}\frac{V^3}{V'^2}d\ln a
\nonumber\\
 &=&
\frac{4}{3m_{pl}^4}\frac{V'^4}{V^4}\int_{\varphi_i}^{\varphi}\frac{V^4}{V'^3}d\varphi\;,
\end{eqnarray}
where we have used the relation $d\ln
a=\frac{8\pi}{m_{pl}^2}\frac{V}{V'}d\varphi$ which can be obtained
from Eq.~(\ref{ava}).

Now we examine a particular phantom inflation model,  the chaotic
inflation with  the phantom potential $V=\frac{1}{2}m^2 \varphi^2$.
Since $|\varphi|$ increases with time $t$ in this potential, if the
slow climb conditions are satisfied at the initial time $t_i$, they
will be forever. This shows that the usual way of exiting from
inflation is unapplicable and this is a problem that needs to be
addressed in phantom inflation scenarios.
Substituting this potential into Eqs.~(\ref{rhobr}, \ref{pbr}) and
imposing the slow climb conditions we find $\rho_{br}\simeq
-p_{br}$.  This shows that the effect of back reaction for the
infrared mode of scalar perturbations acts like a negative
cosmological constant.
 Inserting this chaotic
potential into Eq.~(\ref{Phiva}) yields
\begin{eqnarray}\label{phich}
\langle\Phi^2\rangle=
\frac{2m^2}{9m_{pl}^4}\varphi^2(t)\left(1-\frac{\varphi^6(t_i)}{\varphi^6(t)}\right)\;.
\end{eqnarray}
Substituting the potential  into Eq.~(\ref{ratio}) and using the
above expression, we have
\begin{eqnarray}\label{ratio2}
\frac{\rho_{br}}{\rho_{bg}}\simeq-\frac{2m^2\varphi^2(t)}{3m_{pl}^4}\left[1-\frac{\varphi^6(t_i)}{\varphi^6(t)}\right]
&\simeq& -\frac{2m^2}{3m_{pl}^4}\varphi^2(t)\;.
\end{eqnarray}
Apparently when the ratio becomes of negative unity the back
reaction will terminate the phantom inflation. This requirement
leads to
 \bea\label{vva}
\varphi(t)\geq
\sqrt{\frac{3}{2}}\frac{m_{pl}^2}{m}=\varphi_{cr}\;.
 \eea
 Let us note that one gets approximately the same result for
 estimation of the critical value, $\varphi_{cr}$, by comparing  the effective pressure of
perturbations with the contribution from the phantom potential,
since we have $\rho_{br}\simeq -p_{br}$.
Meanwhile the resolution of flatness problem demands an e-folding of
70 or so. The e-folding number $N$ in the present phantom inflation
scenario can be calculated as follows
\begin{eqnarray}
N=\ln\frac{a}{a_i}=\frac{2\pi}{m_{pl}^2}(\varphi_{end}^2-\varphi_i^2)\simeq\frac{2\pi}{m_{pl}^2}\varphi_{end}^2\;.
\end{eqnarray}
Here $\varphi_{end}$ and $\varphi_i$ denote the value of the
phantom field at the end and the beginning of the inflation
respectively.
 Setting $N\simeq 70$
yields
\begin{eqnarray}\label{varb} \varphi_{end}\simeq
\sqrt{{\frac{35}{\pi}}}m_{pl}\;.
\end{eqnarray}
Thus from Eq.(\ref{vva}) and (\ref{varb}), one finds that if
$m\sim m_{pl}/10$ the phantom inflationary phase can be terminated
by the effects of gravitational back reaction while achieving
efficient inflation e-folds.

Let us note that in this paper only the gravitational back
reaction of long wave-length scalar perturbations is studied,
there are, however, also tensor perturbations (gravitational
waves) in phantom inflation. In normal inflation models, the
effective energy density for the back reaction of long wavelength
tensor perturbations is negative~\cite{Raul} and counteracts the
preexisting energy density. We expect the same for the phantom
inflation.  However, in phantom inflationary scenario,  due to the
quantum instability of vacuum with respect to its decay into
gravitons and phantom particles, the effect of the back reaction
of tensor perturbations may be more significant, but the problem
is more complicated to deal with. We will leave it to our future
investigation. It is worth pointing out, however, that the
amplitude of gravitational waves can not be too large in early
universe because of constraints from nucleosynthesis (see, e.g.,
Ref.~\cite{Davis} and references quoted therein).
\section{Discussions}
In summary, we have calculated the gravitational back-reaction
effects of cosmological perturbations in phantom inflation. Since
the phase space of infrared modes grows rapidly during the
inflation, we only discussed the  the back reaction effects of
long wavelength modes on the background. Our results show that the
effective energy momentum tensor of back reaction has negative
energy density and counteracts the phantom energy.  It is
demonstrated in a particular chaotic phantom inflation model that
when the value of the phantom field, $\varphi$, at the end of the
inflation is larger than a critical value determined by the
necessary e-folds, the back reaction effects could render the
universe to exit phantom inflationary era and enter the radiation
dominated phase.

\begin{acknowledgments}
 This work was supported in part by the National
Natural Science Foundation of China  under Grants No. 10375023 and
No. 10575035, the Program for NCET under Grant No. 04-0784, the
Key Project of Chinese Ministry of Education (No. 205110) and the
Research fund of Hunan Provincial Education Department (No.
04A030)
\end{acknowledgments}

\end{document}